\documentclass[aip,prb,amsmath,amssymb,amstext,citeautoscript,punctuation, reprint]{revtex4-1}
\usepackage{{graphicx,charter, gensymb, float}}
\usepackage{booktabs}
\usepackage[table,xcdraw]{xcolor}
\begin{document}

\sloppy

\title{Accelerated search for new ferroelectric materials}  

\author{Ramon Frey}
\affiliation{Materials Theory, ETH Zurich, Wolfgang-Pauli-Strasse 27, 8093 Zürich, Switzerland}
\author{Bastien F. Grosso}
\affiliation{Materials Theory, ETH Zurich, Wolfgang-Pauli-Strasse 27, 8093 Zürich, Switzerland}
\author{Pascal Fandré}
\affiliation{Materials Theory, ETH Zurich, Wolfgang-Pauli-Strasse 27, 8093 Zürich, Switzerland}
\author{Benjamin Mächler}
\affiliation{Materials Theory, ETH Zurich, Wolfgang-Pauli-Strasse 27, 8093 Zürich, Switzerland}
\author{Nicola A. Spaldin}
\affiliation{Materials Theory, ETH Zurich, Wolfgang-Pauli-Strasse 27, 8093 Zürich, Switzerland}
\author{Aria Mansouri Tehrani}
\affiliation{Materials Theory, ETH Zurich, Wolfgang-Pauli-Strasse 27, 8093 Zürich, Switzerland}
\email{aria.mansouri.t@mat.ethz.ch}

\date{\today}

\begin{abstract}
We report the development of a combined machine-learning and high-throughput density functional theory (DFT) framework to accelerate the search for new ferroelectric materials. The framework can predict potential ferroelectric compounds using only elemental composition as input. A series of machine-learning algorithms initially predict the possible stable and insulating stoichiometries with a polar crystal structure, necessary for ferroelectricity, within a given chemical composition space. A classification model then predicts the point groups of these stoichiometries. A subsequent series of high-throughput DFT calculations finds the lowest energy crystal structure within the point group. As a final step, non-polar parent structures are identified using group theory considerations, and the values of the spontaneous polarization are calculated using DFT. By predicting the crystal structures and the polarization values, this method provides a powerful tool to explore new ferroelectric materials beyond those in existing databases.

\end{abstract}
\maketitle 

\section{Introduction}

% introduce FE and discussion on finding new FE

Ferroelectrics are functional materials that display a switchable spontaneous electric polarization, usually achieved through atomic displacements that connect the ground-state polar crystal structure to a non-polar parent structure. The search for new ferroelectrics is an active area of research due to the multitude of possible applications ranging from capacitors in random access memory \cite{Kohlstedt2005,Ishiwara2012}, to transducers\cite{Cross1987}, and photovoltaics \cite{Paillard2016a,Wang2016}.

Beyond the traditional trial and error attempts, researchers have utilized different strategies and tools in the pursuit of discovering new ferroelectric materials. For example, symmetry and atomic displacement considerations have been used to screen crystallographic databases for new ferroelectrics\cite{Abrahams1988, xu_designing_2017}. The combination of improved computational resources and electronic-structure methods facilitated high-throughput screening of materials and led to new ferroelectric candidates\cite{Bennett2012, garrity_high-throughput_2018, Smidt2020}.
More recently, machine-learning methods have been used to guide searches for high-temperature ferroelectric perovskites\cite{Balachandran2017} and to the identification of ferroelectric photovoltaic perovskites\cite{Lu2019}. The success of applying ML methods to perovskite ferroelectrics motivates our study to construct a generalized framework for accelerating the discovery of new ferroelectrics.

One central task for effectively applying machine-learning algorithms to materials-science problems is developing appropriate descriptors. Descriptors are vector-based numerical representations that uniquely define the material and are usually based on compositional or structural features. It is convenient to use features that rely only on the chemical formula to screen for new compounds so as not to require \textit{a priori} knowledge of the crystal structure. Accordingly, machine-learning models have been developed using compositional features to predict and screen for properties ranging from band gap and ferromagnetism to load-dependent hardness of materials\cite{zhuo_predicting_2018, Long2021accelerating, zhang_finding_2021}. However, since the main factor to be considered in a potential ferroelectric is a polar crystal structure, it is not sufficient to rely on compositional features to search for new ferroelectrics. One avenue to address this issue is first to predict the crystal structure given a particular composition. 

Lately, methods based on density functional theory (DFT), combined with evolutionary algorithms or random sampling of structure spaces, have been developed to predict the ground-state crystal structure of materials\cite{wang_calypso:_2012, glass_uspex-evolutionary_2006, pickard_high-pressure_2006}. However, these methods tend to become computationally expensive due to the vast configuration spaces that need to be assessed for compositions with more than three elements. Thus, machine learning provides an opportunity to accelerate the prediction of crystal structures. One approach to utilizing machine learning for crystal structure prediction is to construct composition-crystal structure correlations\cite{graser_machine_2018}. The main challenge for such an approach is the broad and imbalanced range and distribution of the existing crystal structure types among the more than 10'000 cases in the crystalline databases. Another approach is to create machine learning-based inter-atomic potentials\cite{deringer_machine_2019} for specific phase spaces.  However, fast and accurate prediction of crystal structures remains an open challenge.

A ferroelectric candidate must possess a polar crystal structure, be insulating and be thermodynamically stable. In addition, there should exist a reasonable distortion path between the low-symmetry polar phase and a higher-symmetry non-polar phase. In this work, we develop a framework that consists of a series of machine-learning models in conjunction with high-throughput DFT calculations and group theoretical analysis to screen for new ferroelectric materials requiring only chemical composition as input. We demonstrate the application of our framework by first applying it to chemical spaces that contain known ferroelectric compounds (\textit{e.g.}\,Ba--Ti--O) and then screening for new ferroelectric compounds of quaternary sulfide and selenides. The machine learning models and codes used in this work are provided in the open-source GitLab repository at "https://gitlab.ethz.ch/ramfrey/ml\_fe".

\section{Computational details}
\subsection{Machine learning}

A regression model was constructed to predict the formation energy of the compositions using a random-forest algorithm. DFT formation energies of ternary and quaternary compounds, acquired from the Materials Project database\cite{JAIN20112295,PhysRevB.84.045115}, were used for training. For compositions with more than one entry (polymorphs), only the one with the lowest DFT energy was kept, leading to 65'120 data points. 

A balanced random-forest classifier, as implemented in the imbalanced-learn library \cite{JMLR:v18:16-365}, was used to distinguish between metals and non-metals. For this classifier, the parameter ``sampling strategy'' determines a balanced random-forest classifier, as implemented in the imbalanced-learn library \cite{JMLR:v18:16-365}, was used to distinguish between metals and non-metals. For this classifier, the parameter ``sampling strategy'' determines the way undersampling is performed to deal with the imbalance problem; within this model, it was set to 1 which means that the majority class is undersampled to contain as many entries as the minority class. For training, band gaps were also extracted from the Materials Project in addition to the experimental band gap data compiled in Ref.~\onlinecite{zhuo_predicting_2018}.

Further, a model was developed to classify the compositions into polar or non-polar crystal structures before classifying the point group of any predicted polar compounds. 
The point-group classifier and the band-gap model were constructed using a balanced random forest classifier algorithm. For the polar/non-polar model, the ``sampling strategy'' parameter was again set to 1. The ``sampling strategy'' was within the hyper-parameter optimization for the point-group model. Training data for these two models were extracted from the Pearson crystal structure database (PCD), considering only ternary and quaternary compounds around room temperature and ambient pressure (270--310 K and 1--3 bar). Compounds that did not have a structure prototype assigned, had multiple polymorphs, or contained radioactive elements were discarded. The remaining dataset consisted of 87'153 compounds.

The data were represented by descriptors based on 31 elemental properties (Table\,1 in the supporting information) and their mathematical expressions (average, difference, minimum, maximum, and sum), resulting in 155 features. All models were created using the Scikit-learn library within the python environment\cite{scikit-learn}. For all models, the hyper-parameters, that is the number of trees, maximum tree depth, maximum number of features, and bootstrap, were optimized using a grid search with 10-fold cross-validation with an 80:20 train/test split. For the regression models, the R$^{2}$-score and for the classifiers, the balanced accuracy (obtained using the 
``balanced\_accuracy\_score'' implementation in Scikit-learn) were used as evaluation metrics for the grid search.

\subsection{Density functional theory}

All DFT calculations were performed with the Vienna ab-initio simulation package (VASP) using a plane-wave basis set with projector-augmented-wave pseudopotentials (PAW) and Perdew–Burke–Ernzerhof (PBE) exchange-correlation functionals\cite{kresse_ab_1993, kresse_efficient_1996, kresse_ultrasoft_1999, blochl_projector_1994, perdew_generalized_1996}. To appropriately treat strongly-correlated electronic systems, an on-site effective Hubbard $U_{eff}$ = $U\,-\,J$ correction was applied to the transition-metal $d$ states of compositions containing the elements Co, Cr, Fe, Mn, Mo, Ni, V, and W, according to the standard VASP input sets for Materials project, as implemented in pymatgen\cite{ONG2013314}. 
High-throughput DFT calculations were automated using in-house python scripts and the pymatgen library to prepare and process input and output files. DFT calculations were performed in three subsequent rounds of increasing convergence criteria for efficiency. The criteria for electronic energies and ionic forces were set to 10$^{-4}$\,eV and 0.1\,eV/\AA\, for the first round and then were tightened to 10$^{-6}$\,eV and 0.01\,eV/\AA, respectively. The energy cutoff values were selected, for each compound, based on the maximum default cutoff energy values of the constituent elements that are included in the pseudopotential basis set for the first round and were multiplied by $1.2$ for the second round.
For rounds one and two, k-points were generated automatically using a gamma-centered Monkhorst-Pack scheme, with the number of subdivisions along each reciprocal lattice vector calculated as follows: $max(2,round(20/a_{i}))$, $a_{i}$ being the length of the lattice vector in \AA. For the third round, the standardized input parameters from the Materials Project\cite{JAIN20112295,PhysRevB.84.045115} were set to ensure consistency for the construction of meaningful convex hulls. For sulfides and selenides, one final set of calculations was performed with tighter convergence criteria of 10$^{-8}$\,eV and 0.001\,eV/\AA\, for electronic and ionic parts, respectively, with a plane-wave energy cutoff of 600\,eV. Finally, to confirm the dynamical stability of the predicted crystal structures, phonon calculations were performed using the Phonopy package. \cite{phonopy}

%bader charges were calculated\cite{HENKELMAN2006354}.

\section{Results and Discussions}
\subsection{Workflow}

In Fig.\,\ref{fig1}, we summarize our framework for predicting potential ferroelectric compounds starting from a set of chemical elements. We begin by identifying thermodynamically favorable compositions that are insulating and likely adopt a polar crystal structure using machine-learning models. Next, we use a combination of ML, high throughput DFT, and symmetry analysis to determine the exact crystal structure of the predicted compositions and further downselect the ferroelectric candidates. In the following, we discuss these steps, illustrated by Fig.\,\ref{fig1}, in detail.

\subsubsection{Screening potential ferroelectric compositions}

\begin{figure}[H]
\includegraphics[width=3.6in]{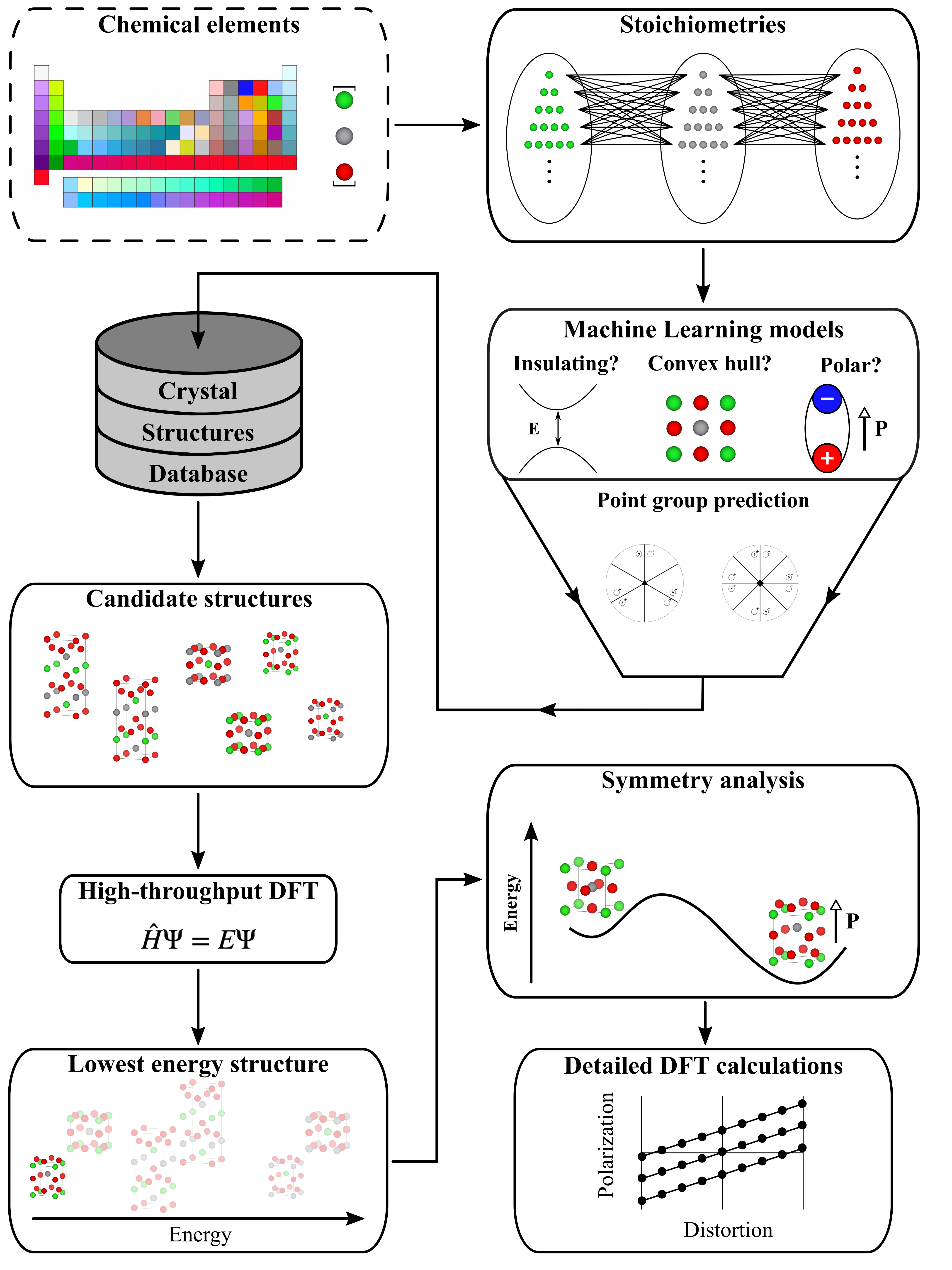}
\centering
\caption{Schematic of the developed workflow. The workflow consists of 9 steps illustrated by 9 panels. First panel: choice of chemical elements. Second panel: generation of the different possible compounds by varying the stoichiometries. Third panel: three machine learning models predict each compound's band gap,  energy above the convex hull, and polarity, then finally the point group. Fourth and fifth panel: according to the stoichiometry and point group predicted, all the possible prototype structures are extracted from the PCD database. Sixth panel: the potential candidate structures are relaxed using high-throughput DFT to identify the lowest energy structure (seventh panel). Finally, symmetry analysis is used to identify the higher-symmetry parent structure (eighth panel), and detailed DFT calculations provide details about the new ferroelectric candidate (ninth panel). }
\label{fig1}
\end{figure}

We start by using a series of ML models to quickly screen composition spaces so that we focus only on viable compositions. As shown in Fig.\,\ref{fig1}, in this part, we develop three ML models to predict compositions with favorable formation energies, insulating behavior, and polar crystal structure.

\noindent Chemical elements are used to construct binary, ternary, and quaternary composition diagrams, which are further evaluated by ML models. Each composition is represented by a vector constructed based on its constituent elemental properties. 
First, a ML model evaluates the thermodynamic stability of the compositions by predicting their formation energies based on Materials Project entries (65'120 DFT calculated energies). The data obtained from the Materials Project are first randomly split into training and test sets with a ratio of 80:20 and are then utilized to construct a ML model using a random-forest regressor. %We chose a random-forest regressor based on initial tests against other algorithms, including support vector regression and multi-layer perception neural network.% 
We achieved high statistical scores of coefficient of determination $R^2 = 0.93$  and root mean squared error $RMSE = 0.32$\,eV by evaluating the random forest classifier on the test set, which included 13'024 compounds. The regression curve that compares the ML and DFT formation energies on the test set is shown in Fig. S1 of the supporting information. Unfortunately, formation energy is not a good indicator of thermodynamic stability, as compounds with negative formation energy might decompose to other competing phases. Therefore, we construct convex hulls based on the predicted formation energies and only consider compositions that are predicted to be within 150\,meV above the convex hull for further evaluation within the framework. The reasons for adding the 150\,meV buffer are twofold. First, it ensures that errors in our prediction do not lead to discarding promising candidates. Second, many compounds in crystalline databases are, in fact, thermodynamically metastable and therefore marginally above the convex hull, though still synthesizable.

\noindent The compositions that are predicted to lie on or near the convex hull are further evaluated for their metallic/insulating behavior, which we take to be the absence or presence of a band gap in the electronic density of states. A classifier is constructed using a balanced random-forest algorithm based on DFT data combined with experimental band gaps (a total of 68'853 entries). Similar to the formation-energy model, the data set is split into training:test sets with a ratio of 80:20. The classifier exhibits an outstanding 87.5\,\% accuracy on the test set. The confusion matrix is also determined to evaluate further the model's ability to correctly differentiate between metals and nonmetals, as shown in Fig.S2a in the supporting information. The diagonal elements are ideally all equal to 1 in a confusion matrix, corresponding to true positives and true negatives. For example, true positives indicate metals correctly predicted to be metallic, while true negatives refer to insulators correctly determined to be insulating. The true positive and the true negative elements of our developed classifier are 0.86 and 0.9, respectively illustrating the high predictive power of the ML model.

\noindent Next, we develop a ML model that takes the compositions selected from the previous steps and determines whether they are likely to adopt a polar or a non-polar crystal structure, see Fig.\,\ref{fig1}. For this model, 87'153 structures are extracted from Pearson's crystal structure database, containing only 5'960 polar crystal structures. Such a significant imbalance in the data can lead to a bias toward predicting compositions as non-polar.
We address this issue by implementing a balanced random-forest algorithm that randomly undersamples the majority class for each tree separately. Using this approach, an accuracy of 83.8\,\% is achieved with a balanced confusion matrix where false positives and true negatives are 0.84 and 0.86, respectively, shown by Fig. S2b in the supporting information. However, it is worth noting that this accuracy is slightly overestimated due to data leakage from training to test set arising from compounds with similar stoichiometries, such as compounds with small dopings and solid solutions. Therefore, we grouped all non-integer stoichiometries to their closest integer equivalent and stratified the train-test splits accordingly to obtain a more conservative statistic. As a result, the accuracy of the classifier drops by 1\%, see Fig. S3a in the supporting information. We note that there is no unequivocal way to decide on which stoichiometries should be grouped, and the choice of grouping is somewhat arbitrary. 

So far, our workflow predicts stable and insulating compositions with polar crystal structures utilizing three successive machine learning models. This screening method significantly reduces the composition space required to search for new ferroelectric materials. 

%However, to calculate the polarisation values and more accurately assess thermodynamic stability and the band gap of the candidates using DFT, we need to know the exact crystal structure.

\subsubsection{Crystal structure prediction}
Next, we use a combined ML/DFT method to predict the crystal structure of the candidates. This approach is more effective than the direct prediction of crystal structures by machine learning because of the large number of classes and imbalanced distribution of the existing data in crystal structure databases.

%One approach to determine the crystal structure of the compositions identified in the previous section is to construct an ML model capable of such prediction directly. However, the large number of classes and the imbalanced distribution of the existing data in crystalline databases render the construction of reliable ML models extremely challenging. Therefore, we combine ML with high throughput DFT to predict the crystal structure as an alternative solution.

\begin{figure}[H]
\includegraphics[width=3.4in]{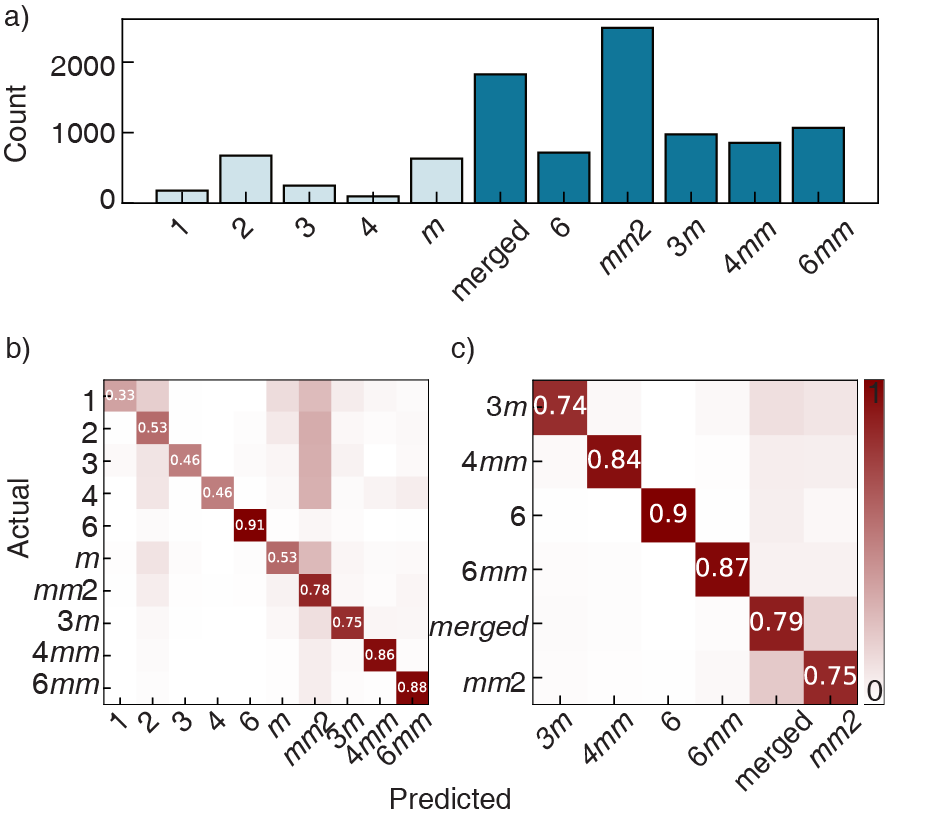}
\centering
\caption{a) Distribution of the point-group symmetries of the compounds in the Pearson Crystal Database (PCD) b) Confusion matrix of the point-group classifier c) Confusion matrix for the point-group classifier when the undersampled classes are merged}
\label{fig2}
\end{figure}

\noindent The first step is to predict the point-group symmetry of the compositions using ML. In total, there are 10 point-group symmetries that constitute the non-centrosymmetric polar crystal classes: $1$\,($C_1$), $2$\,($C_2$), $3$\,($C_3$), $4$\,($C_4$), $6$\,($C_6$), $m$\,($C_S$), $mm$2\,($C_{2v}$), 3$m$\,($C_{33v}$), 4$mm$\,($C_{4v}$), and 6$mm$\,($C_{6v}$) in Hermann-Mauguin (Schoenflies) notation.
%Polar crystal class, also referred to as Pyroelectric, are materials with permanent electric dipole moments. We are, therefore, clearly interested in such materials in the search for ferroelectric compounds. 
Unfortunately, the distribution of the polar compounds in Pearson's crystal structure database is highly imbalanced between these 10 point group symmetries. Fig.\,\ref{fig2}a shows the point group symmetry distribution of polar crystal structures in PCD, illustrating the lack of examples for $1$, $2$, $3$, $4$, and $m$ point groups. Interestingly, a significant percentage, 31.5\,\%, of polar crystal structures in the PCD possess the $mm$2 point group symmetry.
This class imbalance leads to difficulties in utilizing a multi-classifier. Even with undersampling, the random-forest multi-classifier yields low prediction accuracies for the minority classes, as shown by its confusion matrix (Fig.\,\ref{fig2}b). For example, it classifies the compositions correctly for the $1$ point group symmetry only 37\,\%  of the time. To resolve this issue, we merged point groups $1$, $2$, $3$, $4$, and $m$ into a single class called ``merged.'' Utilizing this simplification and further undersampling, an accuracy of 80.2\,\% was reached for the test set with a reasonably balanced confusion matrix, Fig.\,\ref{fig2}c. By grouping all non-integer stoichiometries to their closest integer equivalent and stratifying the train-test splits the accuracy of the classifier drops by 3\%, see Fig. S3b in the supporting information.

\noindent Predicting the point-group symmetries narrows down the possible crystal structure prototypes significantly. Therefore, it is now feasible to use DFT to determine the crystal structures. To generate the structural input for the DFT calculations, we take the candidate stoichiometries combined with their predicted point-group symmetries and determine all the likely crystal structure prototypes they can adopt by comparing them to existing materials in PCD. In addition, all possible configurations of the candidates' elements in each crystal structure prototype are automatically generated. 

\noindent To keep the computational cost low, we perform the DFT calculations in three automated steps, with successively tightened convergence criteria. After each step, the total energies are evaluated, and structures with a larger energy difference than a set cutoff energy are discarded. The cutoff energies are set at 1 and 0.25\,eV for the first and second rounds, respectively. Finally, after the third round, the crystal structure with the lowest calculated DFT total energy is considered to be the crystal structure of that composition.

\subsubsection{Ferroelectric compounds}

A ferroelectric candidate has a non-zero spontaneous polarization switchable by applying an electric field. We now address these points using DFT and group theory starting from the crystal structures predicted in the previous section.

\noindent First, we identify high-symmetry non-polar parent crystal structures using group-super group relationships utilizing the PSEUDO tool from the Bilbao crystallographic server \cite{Capillas}. We automate this process by employing web-scraping methods within the workflow. We then compute the atomic displacements that connect the high-symmetry non-polar and low-symmetry polar structures and take the magnitude of these atomic displacements as a proxy for the strength of the electric field likely needed to reverse the spontaneous polarization. All compounds in which the largest atomic displacement is greater than 1.5\,\AA\ are discarded, as they will likely not have a switchable polarization. The atomic displacements are also utilized to estimate the polarization values by multiplying them by the Bader charges, chosen for an initial estimate since they are readily available at no additional computational cost. Subsequently, the final ferroelectric candidates' electronic structure, crystal structure, dynamical stability, and precise polarization values are evaluated using DFT calculations.

\subsection{Case studies and predictions}

All the ML models within the workflow show high statistical accuracy for predicting potential ferroelectric compounds. Next, we test the workflow by screening composition spaces that contain well-known ternary and quaternary ferroelectric compounds, namely BaTiO$_3$, BiFeO$_3$, and SrBi$_2$Ta$_2$O$_9$, to validate our methodology further. We then re-train all ML models excluding any data with Ba$_a$Ti$_b$O$_c$, Bi$_a$Fe$_b$O$_c$, and Sr$_a$Bi$_b$Ta$_c$O$_d$ compositions from training sets to ensure the models are indeed predictive for these case studies. Below, the workflow implementation is discussed for the case Ba--Ta--O as an example, while the other two cases are included in the supporting information, Fig. S4. 

\begin{figure}[H]
\includegraphics[width=3in]{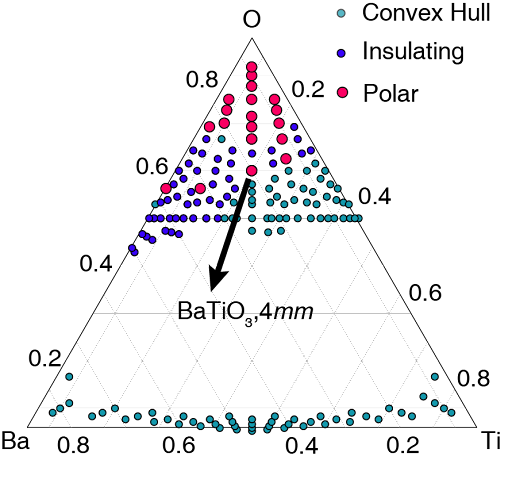}
\centering
\caption{Potential ferroelectric compounds of the Ba--Ti--O composition space identified using ML models that predict proximity to the convex hull, band gap, and polarity of chemical compositions. BaTiO$_3$ is among the selected candidates with predicted point group of 4$mm$.}
\label{fig3}
\end{figure}

Fig.\,\ref{fig3} shows the phase diagram for Ba--Ti--O, which is constructed by first creating a composition grid followed by a series of ML models, as described in the previous section. From 729 initial compositions, our formation energy ML model predicts 172 compositions, highlighted by green circles in Fig.\,\ref{fig3}, to be on or near the convex hull. Next, we predict 79 of these compositions to be insulating (blue circles in Fig.\,\ref{fig3}). Finally, only 19 compositions are predicted to crystallize in a polar crystal structure, shown by red circles. Interestingly, BaTiO$_3$ is among these compositions, demonstrating the ability of ML models to quickly screen composition spaces to focus the search onto viable, functional materials. Our multi-classifier further predicts BaTiO$_3$ to have a 4$mm$ point group. Next, automated DFT calculations are performed to predict the crystal structure of BaTiO$_3$. Our workflow identifies three crystal structure prototypes, \textit{BaNiSn$_3$-type}, \textit{BaTiO$_3$-type}, and \textit{CePtB$_3$-type}, to be compatible with the 4$mm$ point group for BaTiO$_3$. There are two possible sites for Ti in each crystal structure prototype. Therefore, six crystal structures were generated for DFT screening. DFT total energy calculations, performed in three subsequent steps, determine the lowest energy for the case where Ba, Ti, and O adopt the \textit{BaTiO$_3$-type} crystal structure (space group P4$mm$) with each Ti atom coordinated by six oxygens. The calculated total energies for each step are provided in Table 2 in the supporting information. The predicted P4$mm$ crystal structure for barium titanate is the known room-temperature ferroelectric phase of BaTiO$_3$ \cite{Kwei1993}; however, it is not its ground state crystal structure, which has $R3m$ symmetry \cite{Liu2013}. This is expected, as our polar-non-polar and point-group classifiers are trained based on room-temperature experimental data. Similarly, our workflow successfully predicts BiFeO$_3$ ($R3c$) and SrBi$_2$Ta$_2$O$_9$ ($mm2$) as potential ferroelectric compositions, see supporting information Fig. S4. %All the compositions that are predicted to be potential ferroelectrics and their corresponding predicted crystal structure and estimated polarisation values are provided in table SX of the supporting information. 

\begin{table*}[h!]
\centering
\caption{Predicted potential ferroelectric compounds of selected quaternary composition spaces of sulfides and selenides. Space groups are predicted using our combined ML/DFT approach. The energies above the convex hull (E above hull) values are determined using DFT calculations in comparison to the data available in the Materials Project. The band-gap values are determined using DFT calculations at the PBE/PBE+U level. Polarization values are estimated from Bader charges and atomic displacements from the higher symmetry crystal structures, which are determined from group theoretical relationships. Max distortion shows the maximum distance any atom is displaced between the high-symmetry and the polar crystal structure.}
\label{Table}
\resizebox{\textwidth}{!}{%
\begin{tabular}{@{}|l|l|c|c|c|c|c|c|@{}}
\hline
\toprule
\rowcolor[HTML]{C0C0C0} 
\multicolumn{1}{|c|}{\cellcolor[HTML]{C0C0C0}{\color[HTML]{000000} \textbf{\begin{tabular}[c]{@{}c@{}}composition\\  space\end{tabular}}}} & \multicolumn{1}{c|}{\cellcolor[HTML]{C0C0C0}{\color[HTML]{000000} \textbf{composition}}} & {\color[HTML]{000000} \textbf{spacegroup}} & {\color[HTML]{000000} \textbf{\begin{tabular}[c]{@{}c@{}}E above hull\\  {[}eV{]}\end{tabular}}} & {\color[HTML]{000000} \textbf{\begin{tabular}[c]{@{}c@{}}band gap\\  {[}eV{]}\end{tabular}}} & {\color[HTML]{000000} \textbf{\begin{tabular}[c]{@{}c@{}}polarization \\ {[}\textmu C/cm$^2${]}\end{tabular}}} & {\color[HTML]{000000} \textbf{\begin{tabular}[c]{@{}c@{}}high symmetry\\ spacegroup\end{tabular}}} & {\color[HTML]{000000} \textbf{\begin{tabular}[c]{@{}c@{}}max distortion\\  {[}Å{]}\end{tabular}}} \\ \midrule
Pb--Bi--Nb--S & PbBiNbS$_{6}$ & P2$_1$ & 0.09 & 1.2 & 5.5 & P2$_1$/$m$ & 0.95 \\ \midrule
 & PbBi$_2$Nb$_2$S$_9$ & P2$_1$ & 0.17 & 1.12 & 12.0& P2$_1$/$c$ & 1.42 \\
 & Pb$_2$Bi$_3$NbS$_9$ & C$mc$2$_1$ & 0.17 & 0.47 & 3.3 & C$mcm$ & 0.44 \\
 & Pb$_4$BiNbS$_8$ & P$mn$2$_1$ & 0.16 & 0.68 & 5.5 & P$mna$ & 0.93 \\
 & PbBiNbS$_5$ & P1 & 0.13 & 0.86 & 13.1 & P$\bar{1}$ & 1.09 \\
Pb--Bi--Nb--Se & Pb$_2$BiNbSe$_{6}$ & P$na$2$_1$ & 0.11 & 0.7 & 3.6 & P$nma$ & 0.43 \\ \midrule
 & Pb$_2$BiNbSe$_8$ & P1 & 0.18 & 0.93 & 4.9 & P$\bar{1}$ & 1.15 \\
 & PbBiNbSe$_6$ & P$na$2$_1$ & 0.12 & 1.12 & 5.2 & P$nma$ & 1.37 \\
Pb--Bi--Ta--S & PbBiTaS$_{6}$ & P2$_1$ & 0.09 & 1.3 & 8.6 & P2$_1$/$m$ & 1.29 \\ \midrule
 & Pb$_2$BiTaS$_6$ & P$c$ & 0.07 & 1.31 & 0.06 & P2$_1$/$c$ & 0.08 \\
 & BiTaPb$_2$S$_8$ & P1 & 0.12 & 1.43 & 16.5 & P$\bar{1}$ & 1.17 \\
Bi--Ti--Nb--S & BiNbTiS$_{6}$ & P$na$2$_1$ & 0.12 & 0.7 & 3.7 & P$nma$ & 1.01 \\ \midrule
 & BiNbTi$_2$S$_8$ & P1 & 0.20 & 0.44 & 18.3 & P$\bar{1}$ & 1.07 \\
Bi--Ti--Ta--S & BiTaTiS$_{6}$ & P$na$2$_1$ & 0.1 & 0.56 & 16.9 & P$nma$ & 1.29 \\ \midrule
Sr--Bi--Nb--S & SrBiNbS$_{5}$ & P$na$2$_1$ & 0.15 & 0.4 & 7.9 & P$nma$ & 1.43 \\ \midrule
 & BiNbSr$_4$S$_8$ & P$mn$2$_1$ & 0.19 & 1.16 & 6.8 & P$mna$ & 1.31 \\
 & BiNbSrS$_4$ & P$na$2$_1$ & 0.18 & 0.58 & 14.7 & P$nma$ & 0.73 \\
Sr--Bi--Nb--Se & SrBiNbSe$_{5}$ & P$na$2$_1$ & 0.15 & 0.7 & 18.9 & P$nma$ & 1.22 \\ \midrule
Sr--Bi--Ta--Se & SrBiTaSe$_{5}$ & P$na$2$_1$ & 0.12 & 1 & 5.3 & P$nma$ & 0.85 \\ \bottomrule
\hline
\end{tabular}%
}
\end{table*}

Next, given the success of our workflow in correctly identifying known ferroelectric compositions and their crystal structures, we apply it to predict new ferroelectric candidates. The workflow is applied to the composition spaces of 14 quaternary sulfides and selenides, selected based on their similarities to known oxide ferroelectrics\cite{Lines/Glass:Book}. Initially, the ML part of the workflow is utilized to screen 6'561 stoichiometries for each composition space by predicting their convex hull, band gaps, and point groups. On average, 73 compositions (11 if only cases with the most common oxidation states of the constituent elements are considered) are recommended by ML models as potential ferroelectrics, significantly reducing the search space. The ML screening takes less than one hour for each composition space (6'561 stoichiometries) when performed on a typical workstation (Intel Core i7-3770). Therefore, this implementation is ideal for screening numerous compositions. Subsequently, we run all the ML-selected compositions (163 in total) through the rest of the workflow. On average, nine crystal structures were generated, and their DFT energies were elucidated to determine these compositions' crystal structures.  

Finally, our workflow predicts eight new compounds as potential ferroelectrics, listed in Table 1. Note that it did not predict any potential ferroelectric compounds for Cd--Cr--Nb--Se, Cd--Fe--Nb--S, Cd--Fe--Nb--Se, Bi--Ti--Ta--Se, Pb--Bi--Ta--Se and Bi-Ti-Nb-Se composition spaces. In addition to the selected compositions, Table 1 also lists, for each compound, the energy above the convex hull, the DFT-calculated band gap, the estimated polarization values, and the space group of the corresponding high-symmetry crystal structure, which we use for our polarization calculations and to determine the maximum distortion between the two structures. Interestingly, most candidates are predicted to crystallize in the P$na$2$_1$ space group. For example, one of the candidates, SrBiNbSe$_6$ (P$na$2$_1$), is close to the convex hull (0.15\,eV above the hull), is insulating and has a sizable polarization, estimated from the Bader charges. Indeed, our polarization calculations using the Berry Phase method show an even higher calculated polariation value of 31.1 \textmu C/cm$^2$ (supporting information Fig. S5b). Furthermore, beyond the energetic considerations, our calculated phonon density of states (supporting information Fig. S5a) shows that this compound is dynamically stable as indicated by the absence of any imaginary phonon mode. Therefore, SrBiNbSe$_6$ is a promising compound for further experimental and computational exploration as a potential new ferroelectric.

\section{Conclusions}

We developed a workflow that combines machine learning and high-throughput DFT to screen for new ferroelectric materials. The workflow first uses a set of screening criteria, predicted using ML models, to quickly focus the search on viable ferroelectric compositions, followed by prediction of crystal structures and further down-selection of the candidates with the assistance of DFT. Ferroelectricity is connected with the crystal structure of a material, and therefore, an important component of our workflow was our approach for predicting the crystal structure of any polar compound. We validated our methodology by re-discovering known ferroelectric compounds, such as BaTiO$_3$. Moreover, we demonstrated the application of our workflow by predicting new ferroelectric compounds of sulfides and selenides. We hope our work motivates attempts to synthesize and characterize these new potential ferroelectrics. Finally, we note that our method can be applied to any composition space and therefore provides a versatile tool to accelerate the development of new ferroelectric materials by focusing experimental and computational efforts on likely viable candidates.

\section{Acknowledgments}

This research was funded by the European Research Council (ERC) under the European Union’s Horizon 2020 research and innovation program project HERO grant (No. 810451) and the Körber foundation. Calculations were performed at the Swiss National Supercomputing Centre (CSCS) under project IDs s889 and eth3 and on the EULER cluster of ETH Zürich.

\bibliography{ML_FE_Paper, ML_FE_libraries, Nicola}

\end{document}